\documentclass{article}
\usepackage{spconf,amsmath,graphicx,hyperref}

\usepackage{cite}
\usepackage{amsmath,amssymb,amsfonts}
\usepackage{algorithmic}
\usepackage{graphicx}
\usepackage{textcomp}
\usepackage{xcolor}

\usepackage{multirow}

\usepackage{subfigure}

\usepackage{comment}

\usepackage{hyperref}
\hypersetup{hidelinks,
	colorlinks=true,
	allcolors=black,
	pdfstartview=Fit,
	breaklinks=true}

 \usepackage{booktabs}
    
\def\BibTeX{{\rm B\kern-.05em{\sc i\kern-.025em b}\kern-.08em
    T\kern-.1667em\lower.7ex\hbox{E}\kern-.125emX}}

\usepackage[T1]{fontenc}

\usepackage{bbding}

\usepackage{CJKutf8}

\title{AISHELL6-whisper: A Chinese Mandarin Audio-visual WHISPER SPEECH  Dataset with  Speech Recognition Baselines}
\name{Cancan Li$^{1,2}$, Fei Su$^{1,2}$, Juan Liu$^{2,1\dagger}$, Hui Bu$^{4}$, Yulong Wan$^{5}$, Hongbin Suo$^{5}$, Ming Li$^{2,3\dagger}$}
\address{$^{1}$School of Computer Science, Wuhan University, Wuhan, China\\$^{2}$School of Artificial Intelligence,  Wuhan University, Wuhan, China\\$^{3}$Suzhou Municipal Key Laboratory of Multimodal Intelligent Systems, Digital Innovation Research Center, \\Duke Kunshan University, Kunshan, China\\$^{4}$Beijing AISHELL Technology Co., Ltd., Beijing, China\\$^{5}$AI Center, OPPO, Beijing, China}

\begin{document}
\ninept
\maketitle

\footnotetext{$^{\dagger}$ Corresponding author: Juan Liu, Ming Li}

\begin{abstract}
Whisper speech recognition is crucial not only for ensuring privacy in sensitive communications but also for providing a critical communication bridge for patients under vocal restraint and enabling discrete interaction in noise-sensitive environments.
The development of Chinese mandarin audio-visual whisper speech recognition  is hindered by the lack of large-scale datasets. We present AISHELL6-Whisper, a large-scale open-source audio-visual whisper speech dataset, featuring 30 hours each of whisper speech and parallel normal speech, with synchronized frontal facial videos. Moreover, we propose an audio-visual speech recognition (AVSR) baseline based on the Whisper-Flamingo  framework, which integrates a parallel training strategy to align embeddings across speech types, and employs  a projection layer  to  adapt to whisper speech's spectral properties. The model achieves a Character Error Rate (CER) of 4.13\% for whisper speech and 1.11\% for normal speech in the test set of our dataset, and establishes new state-of-the-art  results on the wTIMIT benchmark. The dataset and the AVSR baseline codes are open-sourced at \href{https://zutm.github.io/AISHELL6-Whisper}{https://zutm.github.io/AISHELL6-Whisper}.
\end{abstract}


\section{Introduction}
Whisper speech typically occurs in private conversations under quite and public acoustic environments or in pathological conditions, such as vocal cord damage \cite{farhadipour2024leveraging, lin2023improving, chang2021end}. Automatic whisper speech recognition is valuable not only for enabling discreet  communication in public settings but also for its potential medical applications. Unlike normal speech, whisper speech is produced without vocal cord vibration \cite{jovivcic2008acoustic}, resulting in the absence of fundamental frequency and generally lower energy \cite{ghaffarzadegan2016generative,ito2005analysis}. These characteristics make the accurate recognition  of whisper speech particularly challenging \cite{grozdic2017whispered}.

When collecting the whisper speech using the top microphone of the cell phone, the front-facing camera can also simultaneously capture the lip movement which can be quite useful for speech recognition. However, audio-visual whisper speech recognition remains relatively underexplored, largely due to the scarcity of dedicated datasets. For Chinese Mandarin, only a few small-scale whisper speech corpora are publicly available. The iWhisper-Mandarin dataset \cite{lee2014whispered} was the first released resource, containing approximately 25 hours of whisper and normal recordings for automatic speech recognition (ASR). In \cite{zhou2019audio}, the first Chinese audio-visual whisper dataset  comprising approximately 2.44 hours of recordings from 10 speakers is introduced.



To advance whisper speech recognition, early models \cite{ghaffarzadegan2016generative,grozdic2017whispered} utilized pseudo-whisper data based on Gaussian Mixture Model-Hidden Markov Models (GMM-HMM) to improve performance, while another study \cite{ghaffarzadegan2017deep} employed pseudo-whisper data based on Deep Neural Networks-Hidden Markov Models (DNN-HMM). Chang et al. \cite{chang2021end} proposed an end-to-end (E2E) whisper recognition system, incorporating a frequency-weighted SpecAugment method and a frequency-divided Convolutional Neural Networks (CNN) extractor. Subsequently, Lin et al. \cite{lin2023improving} enhanced E2E performance through pseudo-whisper data augmentation, while Farhadipour et al. \cite{farhadipour2024leveraging} fine-tuned a pre-trained WavLM model \cite{chen2022wavlm} for both whisper and dialect whisper speech recognition. These studies have demonstrated the potential of leveraging deep learning to improve whisper recognition. However, the model structures used in these works still lag behind the performance of advanced speech recognition models published in recent years. Additionally, the effectiveness of training with pseudo-whisper data requires further verification, as it differs from real whisper data and may confuse the model.

In this work, we introduce the AISHELL6-Whisper dataset  to address this gap. It is the largest publicly available audio-visual whisper speech dataset, containing approximately 30 hours each of high-quality whisper speech and parallel normal speech, with synchronized frontal facial videos.

We also implemented Whisper-Flamingo \cite{rouditchenko2024whisper} as the baseline of  our audio-visual whisper speech recognition systems. This model incorporates visual features from AV-HuBERT \cite{shi2022learning} into the OpenAI Whisper speech recognition and translation framework \cite{radford2023robust} through gated cross-attention \cite{alayrac2022flamingo}, achieving state-of-the-art performance in audio-visual speech recognition on the LRS3 \cite{afouras2018lrs3} dataset. We employ a parallel training strategy to align whisper speech embeddings with normal speech embeddings together and apply an additional projection layer on top of the whisper speech features to   further improve the recognition accuracy of whisper speech.
\section{Dataset}
\subsection{Data Collection}
\begin{table}[t]
  \begin{center}
    \caption{Statistics for the Train, Valid and Test sets of AISHELL6-Whisper. 'N' indicates normal speech and 'W' indicates whisper speech.}
    \resizebox{\columnwidth}{!}{
	\label{datasetsplit}
    \begin{tabular}{cccccc}
    \toprule
       \textbf{Set} &\textbf{ Video} &\textbf{ Num of Spk}&\textbf{Type }&\textbf{Time (hrs) }  & \textbf{ Utterances} \\
      \midrule
      \multirow{4}{*}{Train}&\multirow{2}{*}{Yes}&\multirow{2}{*}{82}  &N &  14.8339& 10012\\
                       && &W &14.8359 &9971 \\
                       
    &\multirow{2}{*}{No}&\multirow{2}{*}{28}  &N & 4.8887 & 3335\\
                       && &W & 5.0526 & 3332\\
                       \hline
      \multirow{4}{*}{Valid}&\multirow{2}{*}{Yes}&\multirow{2}{*}{19}  &N &3.3252 &2331\\
                            && &W & 3.3983 & 2304\\
   &\multirow{2}{*}{No}&\multirow{2}{*}{10}  &N & 1.7310&1190\\
                             && &W &1.8038 & 1191\\
                             \hline
      \multirow{4}{*}{Test}&\multirow{2}{*}{Yes}&\multirow{2}{*}{20}  &N &3.4126 &2422\\
                           && &W &3.3601 &2351 \\
    &\multirow{2}{*}{No}&\multirow{2}{*}{8}  &N &1.3477 &938\\
                           && &W &1.3021 &938 \\
       \toprule
    \end{tabular}
    }
  \end{center}
\end{table}

The AISHELL6-Whisper corpus was collected in a controlled studio environment, containing parallel recordings of whisper and normal speech. 
It comprises 167 speakers, each reading approximately 10-20 minutes of poetry texts without any overlap in content.
Among them, 121 participants were recorded using both a high-fidelity microphone and a synchronized RGBD camera, while the remaining 46 participants only recorded the audio signals. Audio was captured with a single-channel high-fidelity microphone (Neumann U87) at a 48 kHz sampling rate, with a background noise level of less than 20 dB. The microphone was positioned below the speaker's chin  to ensure sound quality without obscuring the speaker's lip movement. Video recordings were captured using a RGBD camera placed one meter directly in front of the speaker, with a resolution of 1280×720 at 25 fps.

We divided the dataset into training, validation, and test subsets using an approximate 4:1:1 ratio, ensuring a balanced distribution of age and gender across splits. There are no speaker overlaps between these three subsets. The dataset statistics are summarized in Table \ref{datasetsplit}.

A comparison with other existing whisper speech datasets is provided in Table \ref{comparison}. Our dataset features  relatively larger size, with parallel recording setup and unique multimodal design.
\begin{table}[t]
  \begin{center}
    \caption{Comparison with existing whisper speech datasets. "Parallel" indicates the availability of both whisper and normal speech recordings, "Time (hrs)" refers to the duration of whisper speech.}
    \resizebox{\columnwidth}{!}{
	\label{comparison}
    \begin{tabular}{cccccc}
    \toprule
      \textbf{Dataset} & \textbf{Type} & \textbf{Language} & \textbf{Time(hrs)} & \textbf{Parallel} & \textbf{Num of Spk} \\
\midrule
UTVE-I \cite{zhang2010whisper}& ASR & English &  \textless 1 & \Checkmark & 12  \\
UTVE-II \cite{ghaffarzadegan2014ut}& ASR & English &   1 & - & 112 \\

AVWD \cite{zhou2019audio} & AVSR & Chinese & \textless 2.44 & \Checkmark & 10  \\

CHAINs \cite{cummins2001chains}& ASR & English &  \textless 3 & \Checkmark & 36 \\

Whi-Spe \cite{grozdic2012application} & ASR & Serbian &  \textless 5 & \Checkmark & 10\\

AV-Whisper \cite{tran2013audiovisual} & AVSR & English &  \textless 10 & \Checkmark & 11  \\

CIAIR \cite{kawaguchi2002ciair} & AVSR & Japanese & 15 & \Checkmark & 123  \\

iWhisper-Mandarin \cite{lee2014whispered}& ASR & Chinese & 15 & \Checkmark & 80 \\

wSPIRE \cite{singhal2021wspire} & ASR & English & 18 & \Checkmark & 88  \\

wTIMIT \cite{lim2011computational}& ASR & English & 26 & \Checkmark & 48 \\

\hline
\textbf{AISHELL6-Whisper} & AVSR & Chinese & 30 & \Checkmark & 167  \\
\toprule
\end{tabular}
}
\end{center}
\end{table}

\subsection{Data Processing}
To extract lip movements from the dataset, facial landmarks were extracted using RetinaFace\cite{deng2020retinaface}. Subsequently, lip regions of each frame are cropped following the methodology in \cite{cheng2022dku} using facial landmarks. The mouth center is defined as:
\begin{equation}
    x_{center},y_{center}=\frac{p_2+p_3}{2}
\end{equation}
where $p2$ and $p3$ represent the left mouth corner and the right mouth corner, respectively.
The width of the cropping bounding box is determined by:
\begin{equation}
    width=min\{3.2\times d_{MN},2 \times max\{d_{MN},d_{p1p2}\}\}
\end{equation}
where 
$d_{p1p2}$ denotes the distance between $p2$ and the nose point $p1$, and $d_{MN}$ is the Euclidean distance between $p1$ and the mouth center. Lip movement videos were generated by extracting (96,96) lip regions.


Since speakers may occasionally make minor errors while reading texts, whisper speech may exhibit slight differences from normal speech when reading the same sentence. To address these variations, we calculate the textual similarity for each whisper segment against all normal speech segments from the same speaker
using the 
SequenceMatcher\footnote{\url{https://docs.python.org/3/library/difflib.html}} algorithm. 
The normal speech segment from the same speaker and with the highest similarity score is selected as the matching pair.

\begin{figure}[t]
\centering
\subfigure[Mel spectrogram of a normal speech.]{
\includegraphics[width=0.46\linewidth]{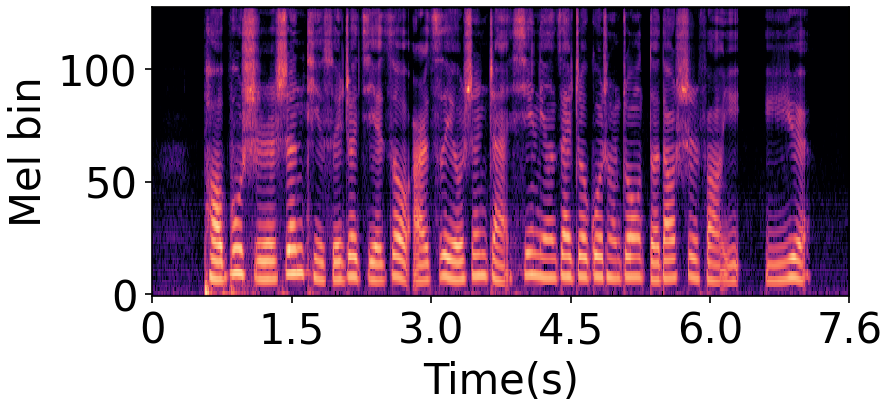}}
\hspace{0.01\linewidth}
\subfigure[Lip movement processing of the normal speech]{
\includegraphics[width=0.46\linewidth]{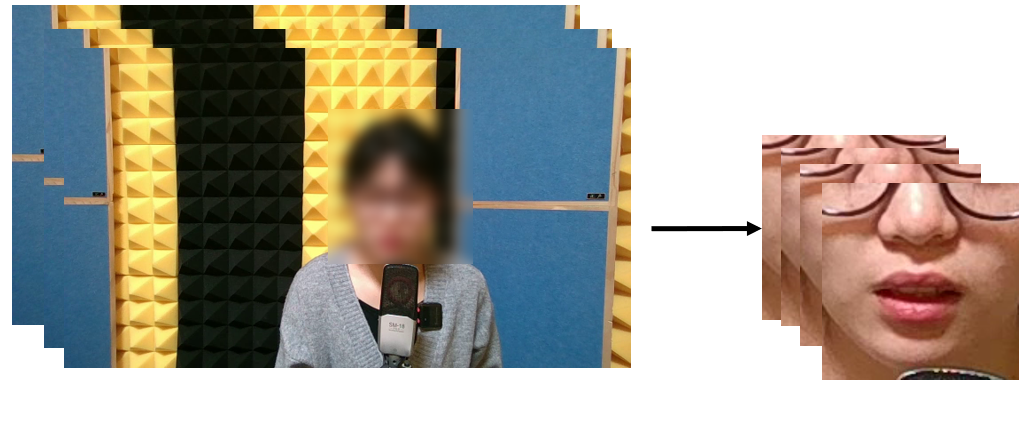}
}
\hspace{0.01\linewidth}
\subfigure[Mel spectrogram of the corresponding whisper speech.]{
\includegraphics[width=0.46\linewidth]{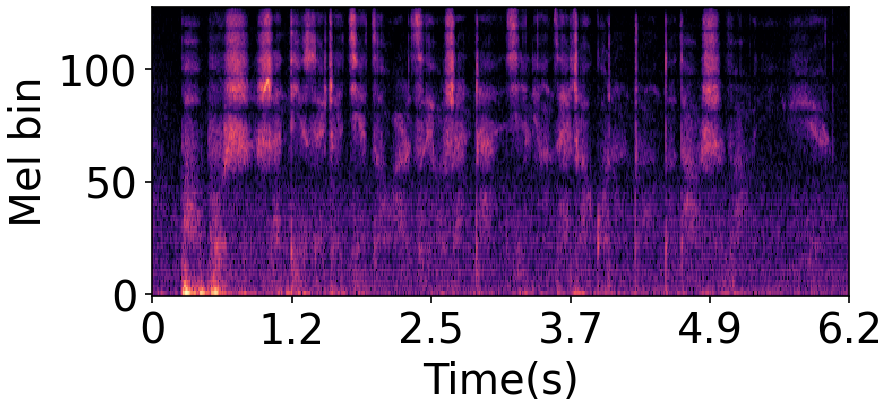}}
\hspace{0.01\linewidth}
\subfigure[Lip movement processing of the whisper speech]{
\includegraphics[width=0.46\linewidth]{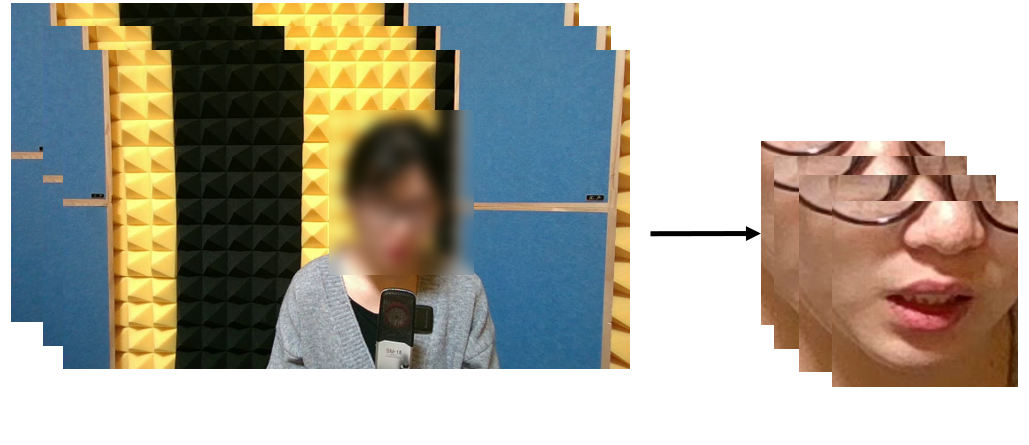}
}

\caption{Data view of AISHELL6-Whisper  with speaker ID “0120” and utterance "\begin{CJK*}{UTF8}{gbsn}
跑步时身体随着节奏而自由伸展心灵在这过程中得到释放与愉悦 
\end{CJK*}".}
\label{view}
\end{figure}

Fig.\ref{view} provides an example of the processed audio and video data, illustrating mel spectrograms for normal speech and the corresponding whisper speech, along with visualized lip movements corresponding to the audio.
The significant spectral differences observed in the  speech spectrogram-particularly the absence of fundamental frequency-demonstrate the acoustic challenges that traditional speech recognition systems face when processing whisper speech.

\begin{figure*}
    \centering
    \includegraphics[width=1.0\linewidth]{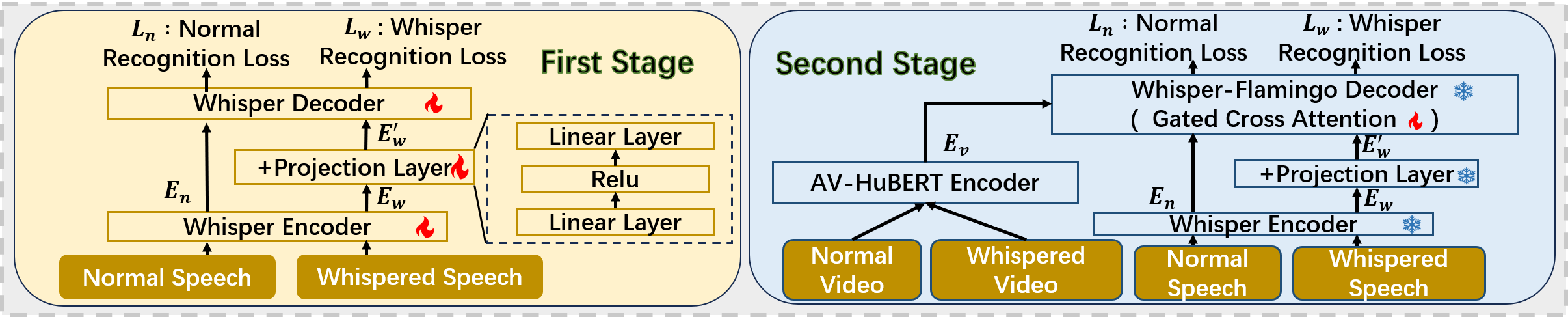}
    \caption{The architecture of our proposed baseline model. In the first stage, paired utterances of whisper and normal speech are simultaneously processed  through a shared Whisper encoder, with a projection layer module refining the whisper embedding. Both embeddings are fed into the decoder to calculate the cross-entropy loss between the model’s predicted transcripts and the ground-truth tokens independently. The second stage follows  the Whisper-Flamingo \cite{rouditchenko2024whisper} approach, incorporating visual features via gated cross-attention for audio-visual training.}
    \label{model_structure}
\end{figure*}

\section{The Proposed Baseline Model}

\begin{table*}[t]
  \begin{center}
    \caption{Speech recognition performance on the AISHELL6-Whisper testing data. AISHELL6-Whisper training set is used for finetuning and training models.}
    \label{ablation_CER}
    \resizebox{0.8\linewidth}{!}{\begin{tabular}{cccccccc}
    \toprule
      \textbf{Model} & \textbf{Parallel training} & \textbf{Projection layer}  & \textbf{Video} & \textbf{Whisper Speech CER} & \textbf{Normal Speech CER} \\
      \midrule
      Whisper (Large-V3)             & - & - & - & 18.93\% & 3.95\% \\
      +Finetune             & - & - & -& 6.69\%  & 1.62\% \\
      +Parallel training          & \Checkmark  & - & - & 4.53\% & \textbf{0.98}\%\\
      +Projection layer         & \Checkmark & \Checkmark & - & 4.34\% & 1.14\% \\
      \hline
      +Video       & \Checkmark & -  & \Checkmark& 4.21\% & 1.08\%\\
      +Video (Proposed)       & \Checkmark & \Checkmark &  \Checkmark& \textbf{4.13\%} & 1.11\% \\
      \bottomrule
    \end{tabular}
    }
  \end{center}
\end{table*}

\subsection{Model structure}

We train the model in two steps, following  the methodology introduced in Whisper-Flamingo \cite{rouditchenko2024whisper}.

In the first step, the OpenAI Whisper model \cite{radford2023robust} is finetuned on the  audio data of whisper and normal speech. Our approach employs a parallel training strategy that processes paired utterances of whisper and normal speech simultaneously, leveraging their intrinsic relationship to enhance model's performance on both speech types, as described in the first stage of Fig.\ref{model_structure}. The Whisper encoder and decoder process paired utterances simultaneously, optimizing the recognition losses for both whisper and normal speech with the following objective:
\begin{equation}
\mathcal{L} = \mathcal{L}_{w} + \mathcal{L}_{n}
\label{total_loss}
\end{equation}

where $\mathcal{L}_{w}$ and $\mathcal{L}_{n}$ represent the standard cross-entropy losses between the model's predicted transcripts and ground-truth tokens for whisper and normal speech, respectively.

In the second step, the model is further finetuned using audio-visual data, following the structure of the second stage in Fig.\ref{model_structure}. In this stage, a gated cross-attention module is integrated into the beginning of each decoder block in Whisper, resulting in the Whisper-Flamingo Decoder. At this training stage, only the gated cross-attention module is trained, allowing the model to incorporate visual input effectively. Both videos corresponding to the  normal speech and whisper speech are fed into the AV-HuBERT Encoder. For data without video signals, video inputs are set to zero.

\subsection{Projection Layer}
To better capture the information absent in whisper speech compared to normal ones, we add a lightweight projection layer module with a linear structure (Linear → Rectified Linear Unit (ReLU) → Linear)  after the Whisper encoder. The two linear components in the structure maintain identical input and output dimensions.  It serves to refine the whisper speech embeddings, aligning them more closely with the acoustic characteristics of normal speech, thereby effectively bridging the  gap for whisper speech. The refinement process is formulated as follows:

\begin{equation}
E_w^{\prime} = E_w + \text{projection layer}(E_w)
\end{equation}

where $E_w$ represents the original whisper speech embedding from the encoder and $E_w^{\prime}$ is the refined embedding. 

The projection layer is carefully initialized to ensure stable training. The first linear layer employs Kaiming normal initialization \cite{he2015delving}, suitable for the subsequent ReLU  activation function. The final linear layer is initialized to zeros, resulting in an identity transformation at the start of training. This initialization strategy allows the model to start from a stable baseline and gradually learn the necessary transformations.

\section{Experimental setup and results}


\begin{table*}[t]
\centering
\caption{Comparison of WER and CER (\%) in the wTIMIT testing data. "Pre-trained" indicates models fine-tuned using checkpoints pre-trained on AISHELL6-Whisper training set data. "Proposed Strategies" indicates using parallel training and projection layer. "N" indicates normal speech and "W" indicates whisper speech, with "US" stands for US accents and "SG" stands for Singapore accents. Chang et al. \cite{chang2021end} evaluated CER on the entire whisper test set and the normal test set, so we provide results that lie between those of the US and SG sets.}
\label{CER_wtimit}
\resizebox{0.9\linewidth}{!}{
\begin{tabular}{lcc|cccc|cccc}
\toprule
\multirow{2}{*}{\textbf{Model}} & \multirow{2}{*}{\textbf{Pre-trained}} & \multirow{2}{*}{\textbf{Proposed Strategies}} & \multicolumn{4}{c|}{\textbf{WER (\%)}} & \multicolumn{4}{c}{\textbf{CER (\%)}} \\
\cline{4-11}
& & & \textbf{$N_{US}$} & \textbf{$N_{SG}$} & \textbf{$W_{US}$} & \textbf{$W_{SG}$} & \textbf{$N_{US}$} & \textbf{$N_{SG}$} & \textbf{$W_{US}$} & \textbf{$W_{SG}$} \\
\hline
 \cite{chang2021end} & - & - & — & — & — & — & \multicolumn{2}{c}{23.5} & \multicolumn{2}{c}{19.0} \\
 \cite{lin2023improving} & - & - & 26.8 & 38.6 & 30.7 & 49.2 & — & — & — & — \\
\hline
Whisper (Large-V3) & - & - & 5.25 & 8.67 & 8.20 & 19.03 & 1.03 & 2.50 & 2.47 & 8.85 \\
+ Finetune & - & - & 5.58 & 7.46 & 7.95 & 15.79 & 0.93 & 1.86 & 2.37 & 7.53 \\
+ Finetune* &  \Checkmark & - & 5.78 & 7.64 & 8.42 & 13.84 & 0.95 & 1.81 & 2.15 & 5.76 \\
\hline
+ Finetune (Proposed)* &  \Checkmark &  \Checkmark & \textbf{4.50} & \textbf{6.06} & \textbf{6.35} & \textbf{11.63} & \textbf{0.73} & \textbf{1.33} & \textbf{1.85} & \textbf{5.17} \\
\bottomrule
\end{tabular}
}
\end{table*}

\subsection{ Database and Evaluation Metrics}

To measure the model's performance on existing benchmarks, it is evaluated on the wTIMIT corpus \cite{lim2011computational}, which contains 450 phonetically balanced sentences spoken in both normal (wTIMIT-n) and whisper (wTIMIT-w) speech by 48 speakers with US and Singaporean (SG) English accents.  Each speech type, either whisper or normal, provides approximately 26 hours of data.
  The current train-test split of wTIMIT contains a significant overlap in utterances, with the same sentences spoken by different speakers, making it unsuitable  for accurately estimating  recognition accuracy \cite{chang2021end}. 
  Partitioning by speakers only reduced performance by approximately 10\% compared to the case with speaker overlap \cite{chang2021end}, and partitioning by speakers can result in less data available for training \cite{lin2023improving}. Therefore,  prohibiting speaker
 overlap between subsets is unnecessary\cite{chang2021end,lin2023improving}. To ensure a consistent comparison with prior works, the dataset is repartitioned in accordance with \cite{chang2021end,lin2023improving}. 
 The resulting splits consist of 400 sentences for training and 25 sentences each for validation and testing for each speaker.

For evaluation, Word Error Rate (WER) and Character Error Rate (CER) are used for the English wTIMIT corpus, while CER is employed for our Chinese speech recognition task.

\subsection{Experimental Setup}
The language identification parameter is set to "zh" and ”en” for Chinese and English speech recognition tasks, respectively.  

All experiments are based on the pre-trained OpenAI Whisper Large-v3 model\cite{radford2023robust}. A maximum audio length of 320,000 samples (20 seconds) is enforced during both training stages. 
 The audio is resampled to 16 kHz to match the Whisper model's  input specification. All other hyperparameters and configurations are adhered to those established in the original Whisper-Flamingo framework \footnote{\url{https://github.com/roudimit/whisper-flamingo}} 
 for fine-tuning the large Whisper and AV-HuBERT models.


In the first stage, the Whisper model is trained for two epochs. In the second stage, the model is further trained with lip movement videos for four epochs.

\subsection{Experimental results}

\subsubsection{Performance on AISHELL6-Whisper}

Table \ref{ablation_CER} presents the CER in the  AISHELL6-Whisper testing data. The original Whisper model exhibits significantly degraded performance on whisper speech, with a CER of 18.93\%, which stands in stark contrast to the 3.95\% CER achieved on normal speech, highlighting the inherent challenge in whisper speech recognition.

Fine-tuning Whisper on the AISHELL6-Whisper training dataset significantly reduces the CER for whisper speech to 6.69\% and normal speech to 1.62\%. Our proposed parallel training strategy further improves performance across both modalities, lowering the whisper CER to 4.53\% and the normal CER to 0.98\%, by enabling mutual reinforcement and compensating for acoustic differences.

The introduction of a lightweight projection layer, specifically designed for whisper speech, further optimizes performance for the target modality, achieving a CER of 4.34\% on whisper speech. As expected, this specialization leads to a slight performance trade-off in normal speech, with the CER increasing to 1.14\%.

Incorporating video information provides complementary visual cues. For the model without the projection layer, video integration enhances whisper recognition (reducing CER from 4.53\% to 4.21\%) but results in a minor regression in normal speech recognition (increasing CER from 0.98\% to 1.08\%). This suggests a shift in the model’s focus toward leveraging the visual modality. Notably, when combined with the projection layer, video integration results in optimal performance, achieving the lowest whisper CER of 4.13\% while maintaining a low normal CER of 1.11\%. This configuration demonstrates the efficacy of our multimodal approach, offering an optimal balance for recognizing both types of speech. Future studies can explore the usage of the video modality under noisy and low SNR scenarios in which visual cues becomes more important.

\subsubsection{Performance on wTIMIT}
To evaluate the generalizability of the proposed techniques, we conducted experiments on the wTIMIT dataset. The wTIMIT training set's limited size (400 sentences for each speaker) increases the risk of overfitting, which makes it difficult for the model to learn whisper-specific features effectively. To address this and improve learning, we initialized our model using checkpoints pre-trained on AISHELL6-Whisper dataset with parallel training and projection layer  strategies, before fine-tuning on wTIMIT. Since wTIMIT only includes audio data, we only trained on the first stage of our proposed model configuration.

As shown in Table \ref{CER_wtimit}, substantial performance gains are achieved on both whisper speech and normal speech.  
The final model, which is fine-tuned using the proposed parallel training and projection layer strategies, demonstrates statistically superior performance to the directly fine-tuned baseline. Specifically, for US accents,  the WER for whisper speech decreases by 1.85\% compared to the original Whisper Large-v3 model. For Singaporean accents (SG), the WER  for whisper speech is reduced by a remarkable 7.40\%, demonstrating that our model not only performs well on our own  AISHELL6-Whisper test data in Chinese but also exhibits exceptional transferability to other datasets in English.

\section{Conclusion}
This work contributes to the field of whisper speech recognition by introducing  AISHELL6-Whisper, the largest audio-visual whisper speech dataset in Chinese Mandarin, 
comprising  30 hours each of whisper speech and parallel normal speech, with synchronized frontal facial videos.

The proposed audio-visual speech recognition baseline, integrating the parallel training strategy with the projection layer  based on the Whisper-flamingo architecture, demonstrates superior performance with a CER of 4.13\% on whisper speech and 1.11\% for normal speech in our test  set while achieving state-of-the-art results on the public  wTIMIT benchmark.

\bibliographystyle{IEEEbib}
\bibliography{refs}

\end{document}